\documentclass{jfm}
\usepackage{graphicx}
\usepackage{psfrag}
\usepackage{natbib}
\usepackage{amsmath}
\usepackage{amssymb}
\usepackage{color}
\usepackage{soul}
\usepackage{subfigure, wrapfig, psfrag} 
\usepackage{mathtools,dsfont}
\usepackage{bm} 
\usepackage{pifont}

\newcommand{\beq}{\begin{equation} }
\newcommand{\eeq}{\end{equation} } 

\newcommand*{\D} {\text{d}}
\newcommand*{\dt} {\text{d}t}

\newcommand*{\xip} {\xi_\ast}

\newcommand*{\kec} {K}
\newcommand{\keccrit} {\kec_{\mathrm{crit}}}
\newcommand{\kecjN}{{\cal K}(j;\Nblock)}
\newcommand*{\KE} {\kec E}

\newcommand*{\Nblock} {N}
\newcommand*{\Nmax} {N_\mathrm{max}}
\newcommand*{\Ntot} {N_T}

\newcommand{\upd} {\mathrm{d}}
\newcommand{\tast} {t_\ast}
\newcommand{\tec} {t_{cv}}
\newcommand{\Cae} {\mbox{Ca}_e}

\usepackage{ifpdf}
\usepackage{graphics}

\ifpdf
    \DeclareGraphicsExtensions{.png, .jpg, .pdf}
\fi
 
\title{Evaporation effects in elastocapillary aggregation}

\author[A. Hadjittofis, J. R. Lister, K. Singh \& D. Vella]%
{Andreas Hadjittofis$^1$, John R. Lister$^2$, Kiran Singh$^1$\\ and Dominic Vella$^1$
\thanks{Email address for correspondence: dominic.vella@maths.ox.ac.uk}
}
\affiliation{$^{1}${Mathematical Institute, Andrew Wiles Building,Woodstock Road,\\ Oxford, OX2 6GG, UK}\\
$^{2}${Department of Applied Mathematics and Theoretical Physics,\\ Centre for Mathematical Sciences, Wilberforce Road, Cambridge, CB3 0WA, UK}
}
\date{\today}

\begin{document} 

\maketitle

\begin{abstract}
We consider the effect of evaporation on the aggregation of a number of elastic objects due to a liquid's surface tension. In particular, we consider an array of spring--block elements in which the gaps between blocks are filled by thin liquid films that evaporate during the course of an experiment. Using lubrication theory to account for the fluid flow within the gaps, we study the dynamics of aggregation. We find that a non-zero evaporation rate causes the elements to aggregate more quickly and, indeed, to contact within finite time. However, we also show that the number of elements within each cluster decreases as the evaporation rate increases. We explain these results quantitatively by comparison with the corresponding two-body problem and discuss their relevance for controlling pattern formation in elastocapillary systems.
\end{abstract}



\section{Introduction}

\begin{figure}
\centering
\includegraphics[width=12cm]{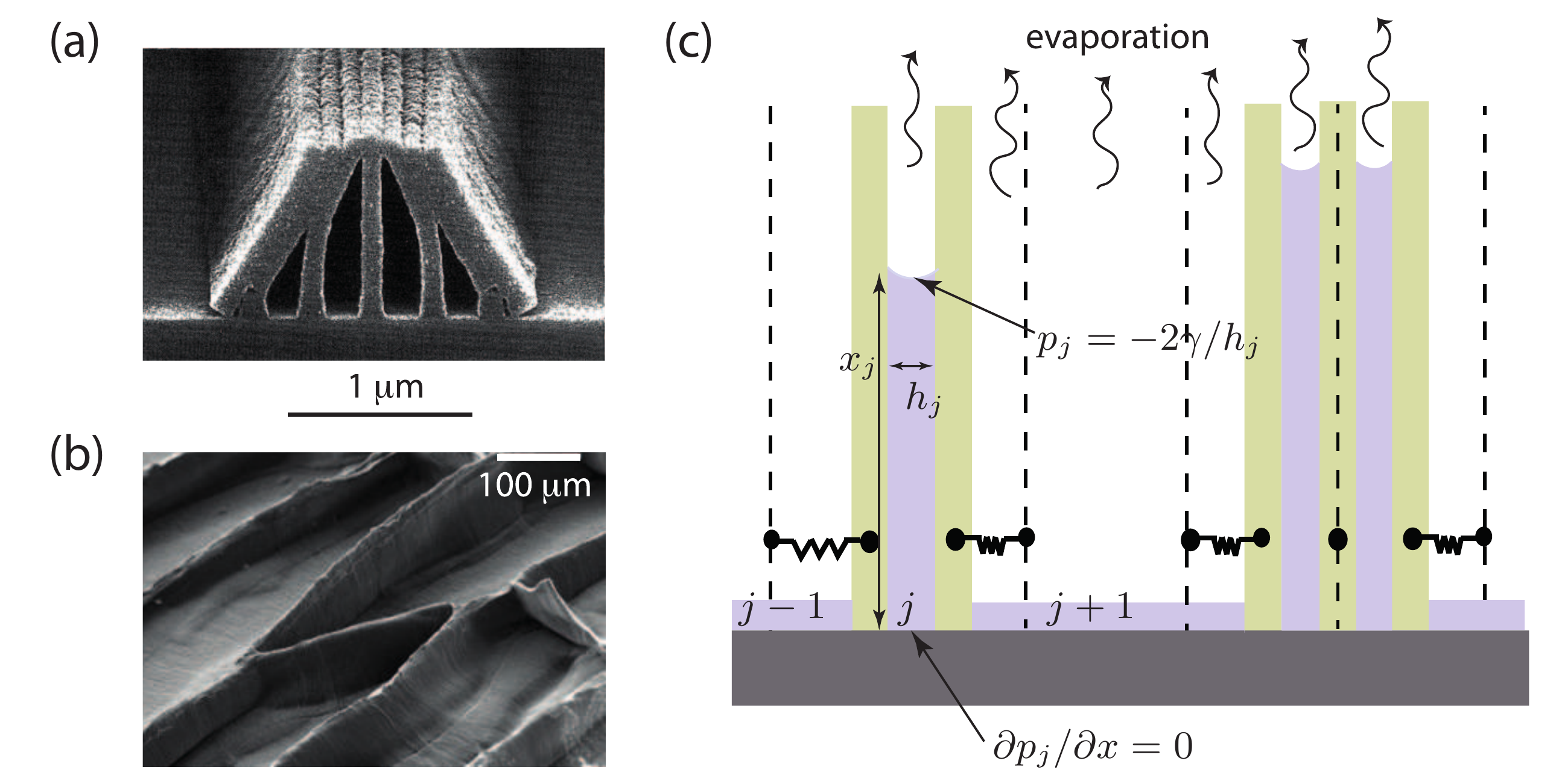}
\caption{Elastocapillary aggregation driven by evaporation of a volatile liquid. (a) Scanning electron micrograph of a pattern created to be part of a MEMS device but damaged by surface tension forces in the rinse step.  \cite[Image reproduced from][Copyright (1993) The Japan Society of Applied Physics.]{Tanaka1993} (b) The formation of cellular foams from the elastocapillary collapse of a forest of carbon nanotubes. \cite[Image reproduced from][Copyright (2004) the National Academy of Sciences]{Chakrapani2004}. (c) The simplified model considered here in which rigid blocks, connected to their initial position by linear springs, replace flexible beams. Lubrication theory in the intervening liquid gaps leads to a second order differential equation for the pressure within each gap, $p_j(x,t)$, which is solved subject to a no-flux condition at $x=0$ and imposed capillary pressure at the meniscus, $x=x_j(t)$, (see Appendix \ref{sec:appDeriv}).}
\label{fig:setup}
\end{figure}

The aggregation of many wet hairs into a series of clumps is familiar from everyday examples including wet paint brushes or eyelashes wetted by tears \cite[see][for example]{Bico2004,Kim2006}. In these scenarios, the surface tension of a liquid acts to minimize the liquid surface area but is resisted by the bending stiffness of the hairs involved. The result is elastocapillary aggregation with a number of clumps of hairs, separated by larger gaps. At a microscopic scale, a similar phenomenon is observed in the manufacture of microelectromechanical systems (MEMS) where long thin elements are etched (using photolithography) and then rinsed \cite*[][]{Tanaka1993}. In the rinse step, the elastic elements are vulnerable to the formation of clumps that can, if the clumps are large enough, lead to fracture (see figure \ref{fig:setup}{\it a}).

While, in many applications, elastocapillary aggregation is an undesirable feature of the process and is best avoided, in other situations it is exploited to form controlled patterns \cite[see][for example]{Chakrapani2004,Pokroy2009}. Indeed, the clumps of carbon nanotubes that form when a nanotube `forest' is wetted and the liquid evaporated (see figure \ref{fig:setup}{\it b}) have sparked considerable interest as a design strategy at microscopic scales \cite[see][for reviews]{deVolder2013a,deVolder2013b}. Despite the importance of other forces (including van der Waals and electrostatic forces) at these small scales, it is clear that capillary forces from the liquid play a vital role: aggregation only happens if the forest has been wetted, and varying the surface tension coefficient changes the properties of the pattern \cite[][]{deVolder2013a,Tanaka1993}.

Most previous theoretical approaches aim to understand the pattern formation in these systems by focussing on energy minimization arguments \cite[see][for example]{Bico2004,Py2007b,Duprat2012,deVolder2011,Jung2014}. This approach neglects the dynamic manner in which the pattern forms and so does not address any features of the pattern that might be controlled dynamically. \cite{Boudaoud2007} developed a toy ballistic aggregation model that gives predictions for the distribution of cluster sizes. However, only recently have detailed dynamic models been developed to capture the interaction between fluid flow and elasticity that occurs in this problem. Several early authors developed  models that couple the elastic deflection with the fluid flow between a pair of elastic beams \cite[][]{Aristoff2011,Duprat2011,Taroni2012}. In particular,  \cite{Taroni2012} showed via their detailed model that even the simple system consisting of a fixed volume of liquid confined between two elastic beams generally exhibits multiple stable equilibria. \cite{Taroni2012} found that which of these equilibria is attained dynamically depends on which is `closer' to the initial condition, rather than which configuration has the lower energy. From this observation, one might expect the final state exhibited by the many-body problem to also be sensitive to the history of deformation, including the initial condition.

Recently, there has been a focus on understanding dynamic aggregation in many-body systems, though to make progress it has been necessary to simplify the elastic problem. \cite*{Gat2013} and \cite*{Singh2014} did this by considering arrays of spring--block elements while \cite{Wei2014} and \cite{Wei2015} considered rigid rods that rotate, resisted by a torsional spring. These models are sufficiently simple that they allow for the development of analytical understanding of the underlying problem while also being able to replicate some of the more complicated behaviour observed in experiments and numerical simulations. For example, \cite{Singh2014} showed that multiple equilibria exist, as in the full problem \cite[][]{Taroni2012}, and that the distribution of cluster size predicted numerically is similar to that observed experimentally by \cite{Boudaoud2007}. Similarly, \cite{Munro2014} demonstrated novel scaling laws for meniscus motion in elastocapillary rise that resemble those found experimentally \cite[][]{Duprat2011}.

Linear stability analysis of the simplified models of dynamic aggregation have all revealed that pairwise aggregation is the mode of aggregation with the largest growth rate \cite[][]{Gat2013,Singh2014,Wei2015}. However, simulations and experiments show that larger clusters form. This apparent discrepancy is believed to be resolved  either as the result of iterated clustering  \cite[in which quadruples form from a pair of pairs etc., as suggested by][]{Gat2013,Wei2015} or by the propagation of a disturbance from a localized perturbation, which leaves behind clusters with a well-defined size \cite[as suggested by][]{Singh2014}.

The models of elastocapillary aggregation discussed above have common features, most notably that the amount of liquid is conserved or is decreased quasi-statically. However, in many applications the liquid is volatile and therefore only wets the elastic elements transiently. Once the liquid has evaporated,  elements can only `stick' together if they were brought sufficiently close together while wet that dry adhesive forces (e.g.~van der Waals forces) come into play. A key question is then: how close to each other do the elastic objects come during evaporation? In this paper we study a reduced model of this dynamic aggregation accounting for the effect of evaporation.

The fluid mechanics of evaporation has seen a great deal of interest in recent years, with particular emphasis being placed on the evaporation of droplets \cite[][]{Cazabat2010}. Much of this work was motivated by the observation that the dark ring left when a drop of coffee evaporates, the so-called ``coffee-ring effect", is caused by the pinning of the contact line and a singularity in the evaporative flux close to the contact line \cite[][]{Deegan1997,Popov2005}. In other cases, the contact line is not pinned so that  the evolution of both the radius and contact angle of the droplet need to be determined dynamically \cite[][]{McHale2005,Stauber2014}. A variety of different evaporation models are available depending on, among other factors, whether the rate of evaporation is limited by diffusion in the vapour phase \cite[][]{Murisic2011}, conduction of heat from a solid boundary \cite[][]{Dunn2009} or convection above the drop \cite[][]{KellyZion2013}. However, even very simple models of evaporation give rise to extremely intricate fluid mechanical problems whose solution can be difficult to distinguish from experimental observations \cite[][]{Oliver2015}.

The coffee-ring effect itself may be viewed as the relatively dilute limit of the drying of a colloidal suspension \cite[][]{Routh2013}. However, at higher concentrations, the densely packed layer of particles left after evaporation cracks just as mud cracks when it dries  \cite[][]{Lecocq2002,Dufresne2006}. Cracks may only propagate in these systems when the elastic energy that is released by fracture is sufficient to pay the associated surface energy penalty, much like a crack in a more classical elastic medium \cite[][]{Dufresne2006}. This problem therefore contains similar ingredients to elastocapillary aggregation, namely elasticity, capillary-induced flow and evaporation.

In this paper, we consider the  elastocapillary aggregation problem accounting for the motion of individual elastic elements within the system, their displacement relative to their initial position, and the fluid flow between elements that controls the dynamics of the process. We shall focus on the effect that the rate of evaporation has on the  pattern that is ultimately formed.

\section{Governing equations}
\label{subsec:model}

To mimic the elasticity of the beams typical in many-body  elastocapillary aggregation, we consider an array of rigid blocks, each connected to their initial position by a linear spring of stiffness $k$ (see figure \ref{fig:setup}{\it c}). The gap between the blocks is considered to be filled with liquid of initial volume $V_0$ (per unit depth into the page); the initial thickness of each gap is $w$.

The surface tension of the liquid, $\gamma$, causes a capillary suction beneath each meniscus, which acts to bring the blocks together. This aggregation is resisted by the springs, mimicking the balance between capillarity and elasticity common to  elastocapillary aggregation in many systems.

To model the dynamics of aggregation, we assume that the gap is relatively thin, $V_0/w^2\gg1$. In this case, lubrication theory can be used to determine the pressure profile $p_j(x)$ within each gap, subject to the boundary conditions that there be no flux through the base ($x=0$) and that the pressure is equal to the capillary suction at the meniscus, $x=x_j$. More details of this calculation  are given in Appendix \ref{sec:appDeriv}; the key result is that the pressure profile within each gap, $p_j(x)$, is determined in terms of the values of $x_j$, $h_j$ and the rate at which the gap is growing/shrinking, $\D h_j/\D t$.   The pressure profile $p_j(x)$ can then be integrated along the wetted length of each side of the gap to give the force on the two neighbouring blocks exerted by the liquid in the $j^\mathrm{th}$ gap:
\beq
F_j=\frac{x_j^3}{h_j^3}\frac{\D h_j}{\D t}+\frac{x_j}{h_j}.
\label{eqn:ND_FP}
\eeq Here the gap thickness and position of the meniscus have been non-dimensionalized by $w$ and $V_0/w$ respectively, force has been non-dimensionalized by $2\gamma V_0/w^2$ and time by the capillary--viscous time
\beq
\tec=\frac{2\mu V_0^2}{w^3\gamma}.
\eeq In \eqref{eqn:ND_FP} the first term represents the lubrication force provided by the liquid in the gap (which resists the motion if $\D h_j/\D t<0$) while the second term represents the net attractive force that results from capillary suction.

Now, each block has two forces acting on it (from the liquid gaps on either side) and the difference of these must be balanced by the net spring force on the block. Subtracting these net forces for the $j^\mathrm{th}$ and $(j-1)^\mathrm{st}$ gaps, we find that
\begin{equation}
2\kec(h_j-1)=F_{j+1}-2F_{j}+F_{j-1}.
\label{eqn:EOM1}
\end{equation} Here the dimensionless spring stiffness
\beq
\kec=\frac{kw^3}{4\gamma V_0}.
\eeq

To close the problem we must also give an equation for the evolution of the position of the meniscus, $x_j$, within the $j^\mathrm{th}$ gap. Previous models have assumed that the total amount of liquid within each gap is conserved, that is $\upd (x_jh_j)/\upd t=0$ \cite[][]{Gat2013,Singh2014}. The dynamics of evaporation are, in general, complicated depending on the presence of convection in the atmosphere, the curvature of the meniscus and evaporation-induced Marangoni effects among others. Furthermore, in the two-dimensional geometry of interest here, the classic approach of considering evaporation limited by steady-state diffusion is ill-posed (essentially because of the logarithmic term in the solution of the two-dimensional Laplace equation). For simplicity, therefore, we assume that evaporation occurs  at a (constant) dimensional rate $e$ per unit surface area. It follows that the volume of liquid within each liquid-filled gap evolves according to
\beq
\frac{\upd }{\upd t}(x_jh_j)=-Eh_j,
\label{eqn:evaplaw}
\eeq where the dimensionless evaporation rate $E$ is defined by
\beq
E=\frac{ew \tec}{V_0}.
\label{eqn:Edefn}
\eeq

Our non-dimensionalization of the problem ensures that a uniform initial condition may be written as $x_j(0)=h_j(0)=1$ for each liquid gap, i.e.~for each $j$. The problem therefore has only two dimensionless parameters: the spring stiffness, $\kec$, and the evaporation rate, $E$. Ultimately, we shall be interested in how the final pattern, particularly the maximum number of blocks in any cluster, depends on these two parameters for systems consisting of a large number of  spring--block elements. However, to gain some understanding of the behaviour of the system in different regions of $(E,\kec)$ parameter space, we first consider the simpler problem of a pair of spring--block elements with a single fluid-filled gap between them. 

\section{An isolated pair: The two-body problem}
\label{subsec:pairs}

For a pair of spring--block elements, there is only a single liquid gap, $j=1$, to consider. In this case $F_0=F_2=0$ so that \eqref{eqn:EOM1} simplifies considerably. Denoting the (only) gap thickness by $h(t)$ and the meniscus position by $x(t)$ we find that
\begin{align}
\frac{x^3}{h^3}\frac{\D h}{\dt}&=-\frac{x}{h}+ \kec(1-h),\label{eqn:EOMpair-h}\\
\frac{\D{x}}{\dt}&=-E-\frac{\D{h}}{\dt}\frac{{x}}{{h}}\label{eqn:EOMpair-x}
\end{align} with initial conditions $x(0)=h(0)=1$.

\subsection{Phase-plane analysis}\label{phase}

The system of equations \eqref{eqn:EOMpair-h}--\eqref{eqn:EOMpair-x} is autonomous so it is natural to examine the behaviour of its phase plane. To facilitate this, we introduce a rescaled gap aspect ratio, $\xi(t)=x(t)/[\kec h(t)]$ and a time-like coordinate $s$ such that $\upd/\upd s=\kec^2h \xi^3\upd/\upd t$ with $s(t=0)=0$. The system \eqref{eqn:EOMpair-h}--\eqref{eqn:EOMpair-x} can then be written
\begin{align}
\frac{\upd h}{\upd s}&=h(1-h-\xi),\label{eqn:EOMpair-hsc}\\
\frac{\upd \xi}{\upd s}&=-\xi\left[\KE\xi^2+2(1-h-\xi)\right].\label{eqn:EOMpair-xi}
\end{align} From \eqref{eqn:EOMpair-hsc}--\eqref{eqn:EOMpair-xi} we see that this (slightly opaque) transformation has eliminated individual occurrences of the  dimensionless parameters $\kec$ and $E$; instead only their product $\KE$ appears. The  dynamics does still depend on the separate values of $K$ and $E$, however, since the initial conditions are now $\xi(0)=1/\kec$, $h(0)=1$. Nevertheless, to understand the phase plane, this rescaling, and the associated reduction of the number of effective parameters, is extremely useful.

\begin{figure}
\centering
\includegraphics[width=13cm]{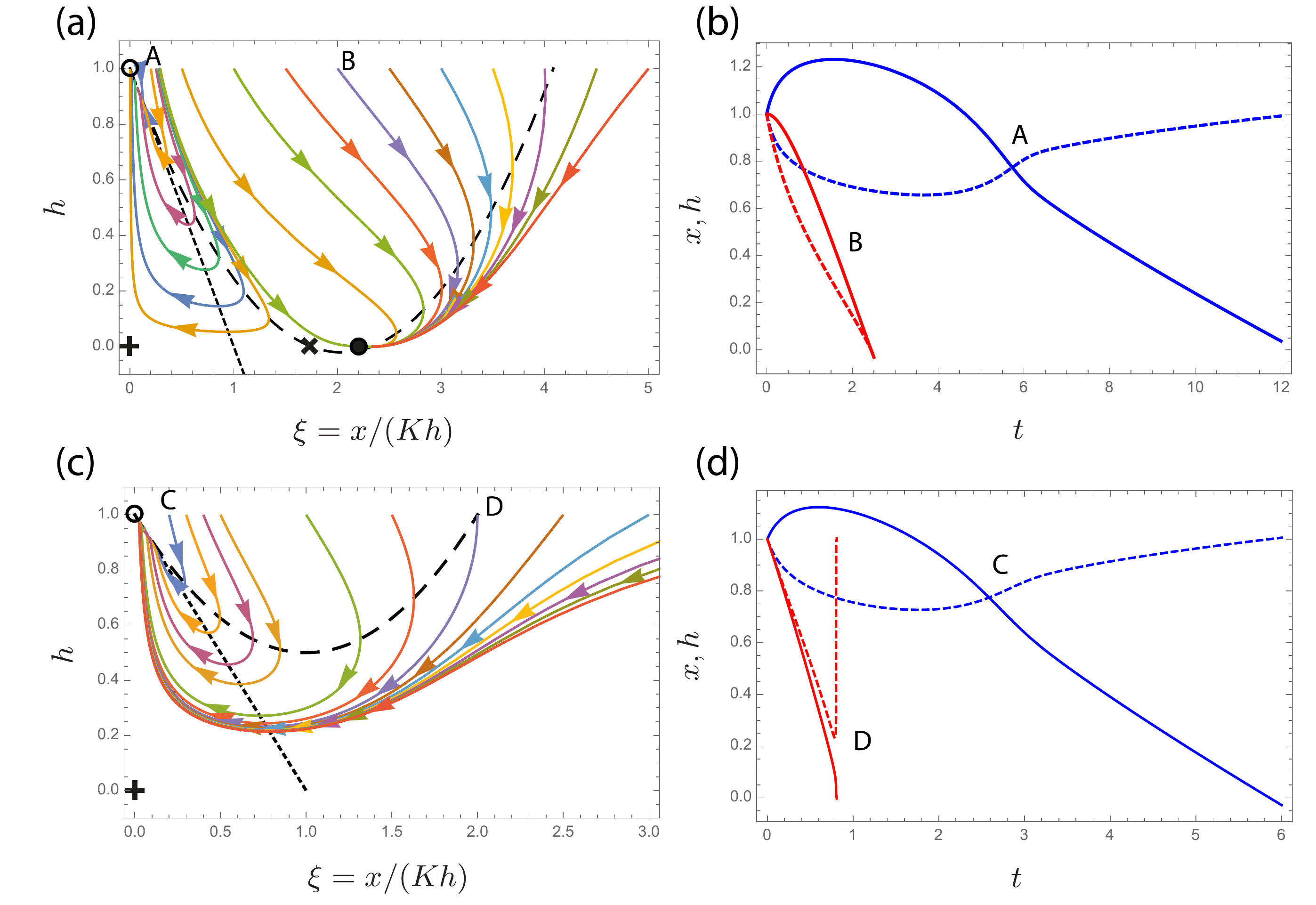}
\caption{ Phase planes and numerically computed trajectories showing the behaviour of the system \eqref{eqn:EOMpair-hsc}--\eqref{eqn:EOMpair-xi} for  $\KE=0.49$ ({\it a} and {\it b}) and $\KE=1$ ({\it c} and {\it d}). In ({\it a}) and ({\it c}) the phase planes show various trajectories (solid curves), corresponding to changing the value of $\kec$ with $\KE$ fixed, with arrows indicating the direction of increasing time-like coordinate $s$ (defined such that $\upd/\upd s=\kec^2h\xi^3\upd/\upd t$). The nullclines $h=1-\xi$ and $h=1-\xi+\KE \xi^2/2$ are shown by the dotted straight line and dashed curve, respectively. For $\KE\leqslant1/2$ there are stable equilibria at $(0,1)$ (separated blocks, indicated by an open circle) and $(\xip,0)$ (stuck blocks, indicated by a filled circle), and two saddle points  at $(0,0)$ (indicated by $+$) and $(\xi_s,0)$ (indicated by $\times$).   For $\KE>1/2$ there is only the equilibrium point at $(0,1)$ (open circle) and the saddle at $(0,0)$ ($+$). ({\it b}) Trajectories of $x(t)$ (solid curves) and $h(t)$ (dashed curves)  for $\kec=5$ (labelled A) and $\kec=0.5$ (labelled B); in each case $\KE=0.49$ and we observe sticking only for sufficiently small $\kec$.  ({\it d}) Trajectories of $x(t)$ (solid curves) and $h(t)$ (dashed curves)  for $\kec=5$ (labelled C) and $\kec=0.5$ (labelled D); in each case $\KE=1$ and sticking is never observed.}
\label{fig:phaseplanes}
\end{figure}

It is clear from \eqref{eqn:EOMpair-hsc}--\eqref{eqn:EOMpair-xi} that the (non-trivial) nullclines in the $(\xi,h)$ plane are $h=1-\xi$ and $h=1-\xi+\KE \xi^2/2$. Combining these with the trivial nullclines $\xi=0,h=0$, we find that if $\KE<1/2$ then the dynamical system \eqref{eqn:EOMpair-hsc}--\eqref{eqn:EOMpair-xi} has two stable fixed points:  $\xi=0, h=1$ (corresponding to two dry blocks in their undeformed, separated configuration) and $\xi=\xip\equiv[1+(1-2\KE)^{1/2}]/\KE$, $h=0$ (corresponding to the two blocks coming into contact as the liquid  between them evaporates). We refer to the fixed point $(\xip,0)$ as the `stuck' configuration since we expect short-ranged forces to maintain contact even after all of the liquid has evaporated. There are also two saddle points for $\KE<1/2$: one at $(\xi_s,0)$ with $\xi_s=[1-(1-2\KE)^{1/2}]/\KE$ and another at $(0,0)$.

A typical phase plane for $\KE<1/2$ is shown in figure \ref{fig:phaseplanes}({\it a}) (with $\KE=0.49$); the evolution of the physical variables for two example trajectories are shown in figure \ref{fig:phaseplanes}({\it b}). The phase plane, figure \ref{fig:phaseplanes}({\it a}),  shows that for $\xi(0)=1/\kec$ sufficiently small, i.e.~for sufficiently stiff springs, the system ultimately tends to the separated fixed point, $(0,1)$. However, for $\xi(0)=1/\kec$ larger than some threshold, i.e.~for sufficiently weak springs, the system instead tends to the stuck fixed point, $(\xip,0)$. This  is illustrated by the evolution of the physical variables shown in figure \ref{fig:phaseplanes}({\it b}). The critical value of $\kec$ at which the transition from separated to sticking  occurs, $\keccrit(\KE)$, can be determined numerically from the stable manifold of the saddle point at $(\xi_s,0)$ (see Appendix \ref{sec:appPP}). For the case $\KE=1/2$ we find that the system tends to the stuck fixed point if $\kec< 3.30$ (i.e.~$E>0.15$) but otherwise becomes separated. In the limit $\KE\ll1$ we find that $\keccrit=\frac{27}4-O(E^{2/3})$, which agrees with the result of \cite{Singh2014} in the absence of evaporation. 

For $\KE>1/2$,  $\xip$ is complex so that the only stable fixed point is at $\xi=0,h=1$; there is also a single saddle point at $(0,0)$. Physically, we deduce that, for sufficiently fast evaporation or sufficiently strong springs, the system must always end up with the  blocks separated and all of the liquid evaporated  --- the blocks are `unstuck'. An example of the trajectories of the system, plotted in $(\xi,h)$ space, is shown in figure \ref{fig:phaseplanes}({\it c}); as expected, this shows that all trajectories (i.e.~regardless of the value of $\kec$) ultimately reach the unstuck state. This is confirmed by the  physical evolution along two example trajectories, as shown in figure \ref{fig:phaseplanes}({\it d}). Note from the case $\kec=0.5$, $E=2$ (labelled `D' in figure \ref{fig:phaseplanes}({\emph d})) that once the blocks start to separate they may do so very rapidly: both evaporation and the widening gap reduce the hydraulic resistance to motion while the total capillary suction in the gap is also dramatically decreased by a decrease in $x$.

\begin{figure}
\centering
\includegraphics[width=13cm]{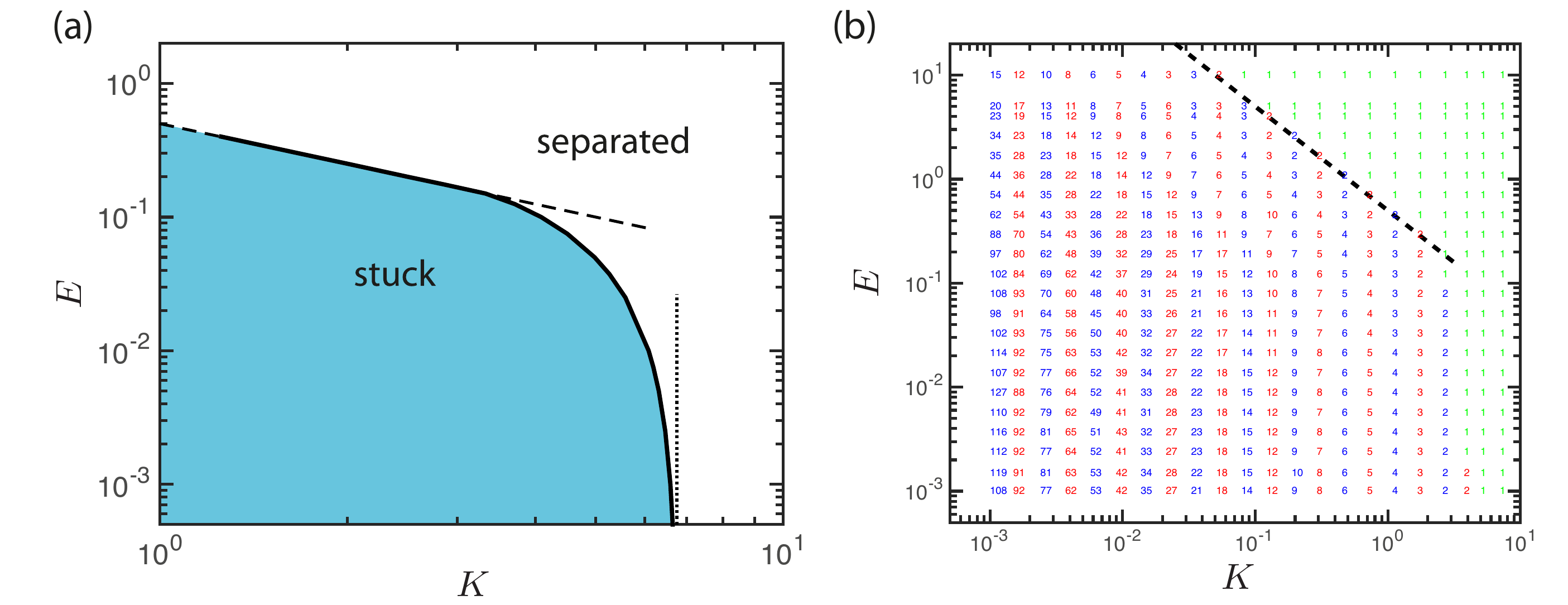}
\caption{Regime diagrams showing the behaviour of the system as functions of the spring stiffness $\kec$ and evaporation rate $E$. (a) The regime diagram for the two-body problem in which only stuck and separated states exist. The solid curve shows the numerically determined boundary between stuck and separated while the results with $E=0$, $\keccrit=27/4$ (dotted vertical line), and  $\keccrit=1/2E$ (dashed line), are also shown. (For $K\lesssim3.30$, the boundary between stuck and separated states is $\KE=1/2$, see text.) (b) Results of numerical simulations with $N=2000$ blocks with randomized initial perturbations showing the largest number of blocks in a cluster, $\Nmax$, at various $(\kec,E)$. Here green shows situations in which no blocks aggregate ultimately while red and blue are used to distinguish results at neighbouring points of $(\kec,E)$ space. The dashed line again shows the result $\keccrit=1/2E$ as a hint that the two-body problem is relevant to understanding the many-body problem.}
\label{fig:2BodyRegimeDiag}
\end{figure}

In summary, our analysis of the phase plane shows that for $\KE>1/2$ the blocks must ultimately tend to the separated state, while for $\KE\leqslant1/2$ the blocks stick if $\kec\leqslant \keccrit(\KE)$ and otherwise separate. This behaviour is summarized in figure \ref{fig:2BodyRegimeDiag}({\it a}), which shows the regions of $(\kec,E)$ parameter space for which the blocks are ultimately stuck or separated. However, we note that in either case the system tends to a state in which all of the liquid has evaporated, $x=0$, whether that be with $h=1$ (the separated state) or with $h=0$, $x/h=\kec\xip$ (the stuck state).

\subsection{Contact occurs in finite time or not at all}

In the limit of no evaporation, $E=0$, \cite{Singh2014} showed that if the springs are sufficiently weak for a pair to stick, then the gap thickness $h$ decreases with time according to $h\sim(3t)^{-1/3}$ for $t\gg1$. In this idealized system, therefore, the blocks do not actually contact one another (though in reality van der Waals attraction or the roughness of the blocks would ensure that contact does eventually occur).

The situation with even a small amount of evaporation, $E>0$, is qualitatively different in that contact occurs within a finite time. To see this we let $h\ll1$, $x/h\approx K\xip$ in \eqref{eqn:EOMpair-h}, yielding
\beq
\frac{\upd h}{\upd t}\approx\frac{1-\xip}{\kec^2\xip^3}=-\frac{E}{2\kec\xip}.
\eeq Crucially, this is constant so that the gap thickness decreases at a constant rate and, hence, the gap must  close and the blocks touch.

Note that in the limit of small evaporation, $E\ll1$, and with $0<\kec<\keccrit(\KE)$ so that sticking occurs, we have $\xip\approx2/\KE$ leading to
\beq
\frac{\upd h}{\upd t}\approx-\frac{E^2}{4}.
\label{eqn:htCollapse}
\eeq We therefore expect the time at which contact occurs to diverge as $E\to0$, consistent with the previous observation that contact does not occur in the absence of evaporation \cite[][]{Singh2014}. This expectation may also be confirmed for the case of no springs, $\kec=0$, where a simple calculation shows that contact occurs at $t=\tast$ with
\beq
\tast=\frac{1}{E}+\frac{2}{E^{3/2}(2-E)^{1/2}}\tan^{-1}\left[\left(\frac{2}{E}-1\right)^{1/2}\right]\sim\frac{\pi}{\sqrt{2}}E^{-3/2}\,\mathrm{ as }\,E\to0.
\label{eqn:K0tast}
\eeq 

\section{The many-body problem}

\subsection{Numerical results}

Our primary interest is in the dynamics of aggregation for a large number of spring--block elements. To gain some insight into the behaviour of these systems, we solved the system of differential equations \eqref{eqn:EOM1}--\eqref{eqn:evaplaw} with up to $\Ntot=2000$ elements. We used symmetry boundary conditions at the edge of the system, i.e.~$F_{-1}=F_1$ and $F_{\Ntot+1}=F_{\Ntot-1}$ \cite[][]{Singh2014}, though similar results were obtained with free edges, $F_0=F_{\Ntot}=0$. Using symmetry boundary conditions has the advantage that the uniform state $h_j=1$ is an equilibrium. The initial condition was then taken to be a small random perturbation to this state, i.e.
\beq
h_j(0)=1+\epsilon {\cal R}_j
\eeq with ${\cal R}_j$ a random number drawn from $N(0,1)$. For the results presented here, we took $\epsilon=10^{-2}$. The initial meniscus position was $x_j(0)=1/h_j(0)$ so that each gap contained the same volume of liquid initially.

Two adjustments to the codes used by \cite{Singh2014} were made to account for the ultimate disappearance of liquid inherent in evaporation. Firstly, the liquid level $x_j(t)$ was prevented from becoming negative by enforcing $x_j(t)\geqslant 10^{-10}$ for all times. Secondly, a short range van-der-Waals interaction was introduced to ensure that once elements come sufficiently close to contact they remain close, but do not pass through one another; this mimics the sticking upon contact that is observed in MEMS devices. Simulations were run until  $\sum_j|\dot{h}_j|<10^{-10}$, so that each gap has reached an equilibrium to within numerical error.

\begin{figure}
\centering
\includegraphics[width=13cm]{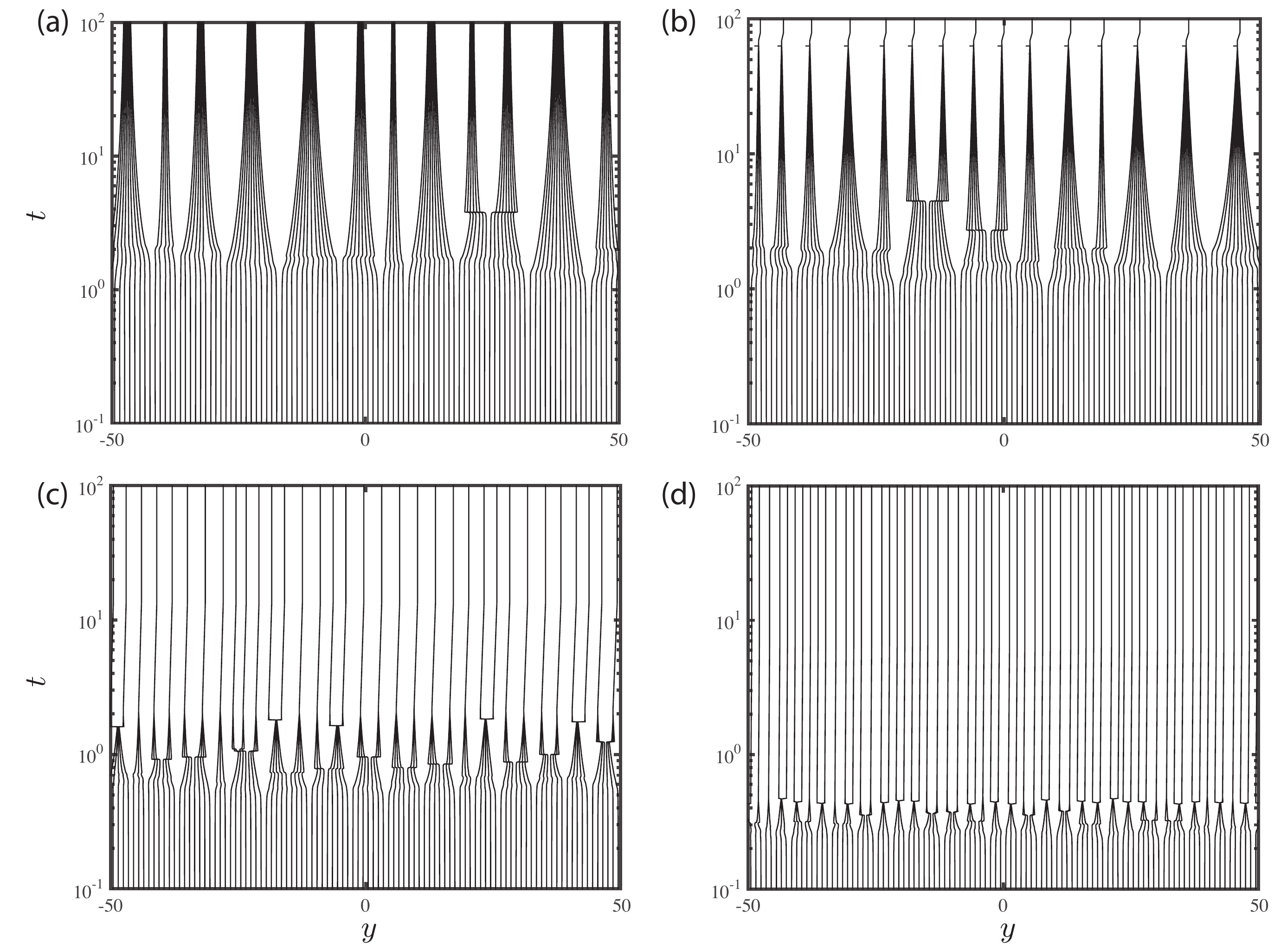}
\caption{Space-time diagrams showing the position of $\approx100$ blocks within a larger simulation of the elastocapillary aggregation of $\Ntot=500$ blocks. The spring stiffness $\kec=0.1$ in each case and the evaporation rate $E$ is varied: (a) $E=0$, (b) $E=0.1$, (c) $E=1$, (d) $E=3$.}
\label{fig:teepee}
\end{figure}

Figure \ref{fig:teepee} shows space-time diagrams for the evolution of around $100$ block positions during aggregation. Results are shown for a range of values of the dimensionless evaporation rate $E$ but with $\kec=0.1$ fixed. Figure \ref{fig:teepee}({\it a}) (in  the absence of evaporation) shows that clusters first form and then shrink down as time increases. 
Introducing a small amount of evaporation (figure \ref{fig:teepee}{\it b}), we see that the typical size of clusters is broadly similar but that the clusters seem to shrink to contact within a finite time. We also observe that there seem to be more scenarios in which clusters start to form but ultimately split --- what we term a `cluster bifurcation'. (These cluster bifurcations are also associated with a `kink' in the space-time diagram because the force balance is radically altered as the cluster splits: the new sub-clusters shift to positions where all of the spring forces are reduced.) Increasing the evaporation rate still further (figures \ref{fig:teepee}{\it c,d}), we see, unsurprisingly, that the clusters that form shrink to contact more quickly than with lower evaporation rates. We also see that there are many more cluster bifurcations and that the final clusters tend to be smaller than those observed with smaller evaporation rates.

\begin{figure}
\centering
\includegraphics[width=11cm]{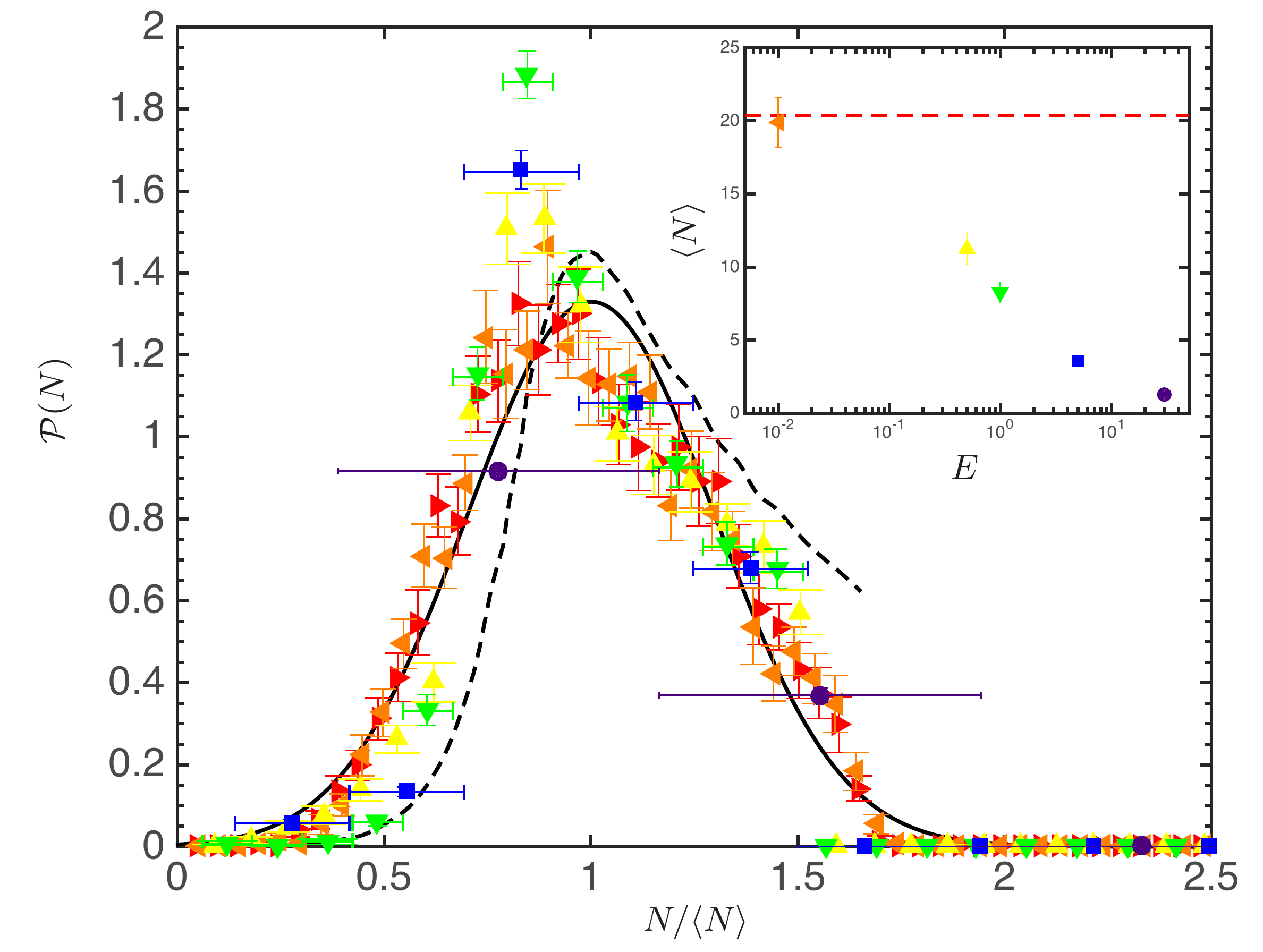}
\caption{The statistical properties of the cluster size $N$ for a range of evaporation rates $E$, with $\kec=0.0137$ fixed. Inset: The mean final cluster size, $\langle N\rangle$, as estimated, for each value of $E$, from 30 numerical experiments with initial conditions $h_j=1+\epsilon{\cal R}_j$, where $\epsilon=10^{-2}$ and ${\cal R}_j$ are random numbers drawn from $N(0,1)$. The vertical error bars represent the standard error. The mean cluster size with no evaporation ($E=0$) is shown by the horizontal dashed line. Main figure: The probability distribution function for the cluster size as found in the numerical experiments. Different evaporation rates are indicated as follows: $E=0$ ($\blacktriangleright$), $E=10^{-2}$ ($\blacktriangleleft$), $E=0.5$ ($\blacktriangle$), $E=1$ ($\blacktriangledown$), $E=5$ ($\blacksquare$) and $E=30$ ($\bullet$). Horizontal error bars show (for large $E$ and small $N$) the discrete nature of the data, while vertical error bars show the standard error. The solid curve shows a normal distribution with mean $1$ and standard deviation $0.3$, as suggested previously \cite[][]{Singh2014}; the dashed curve shows the distribution found by \cite{Boudaoud2007} from a mean-field aggregation model with a single  parameter fitted to match experiments.}
\label{fig:statistics}
\end{figure}

Another perspective on the effect of evaporation is provided by examining the statistics of the cluster size $N$ as $E$ varies with $\kec$ fixed. The inset of figure \ref{fig:statistics} shows that the mean cluster size $\langle N\rangle$ is a decreasing function of the evaporation rate, as should be expected from the earlier observations of  figure \ref{fig:teepee}. Figure \ref{fig:statistics} itself shows the probability distribution function for a range of evaporation rates, $0\leqslant E\leqslant 30$, calculated by averaging the results of thirty runs at each evaporation rate. To allow for a comparison of the shape of the distribution as the evaporation rate varies, we have normalized the $x$-axis in figure \ref{fig:statistics} by  $\langle N\rangle$. In these normalized coordinates, the typical (normalized) width of the distribution does not change noticeably as the evaporation rate increases (i.e.~the width of the true distribution scales with the mean $\langle N\rangle $ and hence does become narrower as $E$ increases and $\langle N\rangle$ decreases). Surprisingly the behaviour in the tails of the distribution seems to follow quite closely that found in the absence of evaporation (this is true even at $E=30$ where only pairs of blocks or single blocks  are observed). However, we note that the peak around $N/\langle N\rangle\approx0.8$ appears to become more pronounced as the evaporation rate $E$ increases up to $E=1$; at the largest values of $E$ the discrete nature of the distribution and the decrease in $\langle N\rangle$  blunts the peak. 

In summary, the simulation results shown in figures \ref{fig:teepee} and \ref{fig:statistics} suggest that the presence of evaporation modifies the dynamics of aggregation in three important ways: (i) any clusters that form shrink down more quickly than would be the case without evaporation, (ii) clusters are more prone to split up, or bifurcate, when evaporation is included, and (iii) the final state generally consists of smaller clusters. The first of these observations recalls what was seen in the two-body problem where contact occurs in finite time with individual blocks moving at a constant speed, see \eqref{eqn:htCollapse}. We shall see that the second and third observations are intimately related.

To better understand the third observation, numerical simulations with  a range of values of $\kec$ and $E$ were performed and  the maximum number of blocks in any cluster noted, see figure \ref{fig:2BodyRegimeDiag}({\it b}). This plot confirms the trend observed in (iii) above. In particular, we  note that for sufficiently large evaporation rates $E$, no clusters are formed (the green points in figure \ref{fig:2BodyRegimeDiag}{\it b}). Comparing the transition between no clustering and clustering observed with many blocks (figure \ref{fig:2BodyRegimeDiag}{\it b}) to the transition between separated and stuck states observed in the two-body problem (figure \ref{fig:2BodyRegimeDiag}{\it a}) we observe a strong  similarity; in particular, the line $\kec E=1/2$ seems to be important in both transitions. Away from this line, the maximum number of elements in a cluster varies  as the evaporation rate $E$ varies (at fixed $\kec$). To try to understand these results, we begin by considering what happens in the limit of no springs.

\subsection{No springs: $\kec=0$}

With no springs, $\kec=0$, the problem simplifies considerably. To see this, we note that \eqref{eqn:ND_FP} then gives
\beq
\label{eqn:K0}
F_{j+1}-2F_j+F_{j-1}=0.
\eeq If the system is unforced at its edges, $F_{0}=F_{\Nblock+1}=0$, then this may be inverted to give
\beq
F_j=\frac{x_j^3}{h_j^3}\frac{\D h_{j}}{\dt}+\frac{x_{j}}{h_{j}}=0
\eeq for $j=1,...,N$. The motion within each gap therefore decouples from that in every other gap and we recover the two-block problem \eqref{eqn:EOMpair-h}--\eqref{eqn:EOMpair-x} with $\kec=0$. We  conclude that in this case all of the blocks will aggregate with contact occurring at $t=\tast$ with $\tast$ given by \eqref{eqn:K0tast}.

\subsection{Finite strength springs $\kec>0$}

Analytical progress is possible in the limit of no springs because the evolution of the liquid within each gap becomes decoupled from that in every other gap. However, with non-zero spring stiffness, the force balance \eqref{eqn:EOM1} must hold; this couples the evolution within the gaps to one another. 

\subsubsection{Analytical results close to contact}

To make analytical progress, we consider the limit of small gap thickness, $h_j\ll1$, so that a cluster of blocks are already close to contact. In this limit additional displacements are small and  the force arising from the springs is approximately constant, simplifying the problem considerably. In particular, when the blocks are close to contact, $h_j\ll1$, the force-balance equation \eqref{eqn:EOM1} may be approximated by
\begin{equation}
-2\kec= F_{j+1}-2F_{j}+F_{j-1}.
\label{eqn:EOM1:smallh}
\end{equation} Physically, this equation expresses the fact that in the small-gap limit, the net hydrodynamic force on a block, $F_{j+1}-F_j$, exceeds that on its neighbour by $2\kec$ since it is approximately one unit further from its equilibrium position.

Given two particular $F_j$ somewhere within the system, \eqref{eqn:EOM1:smallh} may readily be inverted to give an expression for general $F_j$. Motivated by the repeated cluster-bifurcation events seen in figures \ref{fig:teepee}({\it c,d}), we consider whether a cluster of $\Nblock$ spring--block elements (enclosing $\Nblock-1$ liquid-filled gaps) that is close to complete collapse will bifurcate. Labelling the liquid gaps from one side of the cluster, we can  approximate the hydrodynamic force at its edges by zero, i.e.~$F_0=F_{\Nblock}=0$, since the cluster will be comparatively well separated from any neighbouring elements. Inverting \eqref{eqn:EOM1:smallh} subject to these edge conditions, we find that
 \begin{equation}
 F_j = \kec j(\Nblock-j) 
 \label{eqn:Fclust}
  \end{equation} 
%
%
Thus the hydrodynamic force $F_j$ in a gap is given by a constant value $\kecjN\equiv\kec j(\Nblock-j)$ that depends only on the label $j$ of that gap within a particular cluster, and the number of blocks $\Nblock$ in the cluster. (The detailed gap thicknesses do not appear because the displacements of blocks from their equilibrium positions, and the consequent spring forces, are nearly constant when the blocks are close to contact.)

Now, recalling from \eqref{eqn:ND_FP} that the hydrodynamic force  $F_j$ is the sum of lubrication and capillary components, we can write \eqref{eqn:Fclust} together with the equation describing mass conservation as the system  
\begin{align}
\frac{x_j^3}{h_j^3}\frac{\D h_{j}}{\dt}&=-\frac{x_{j}}{h_{j}}+ \kecjN
 \label{eqn:manybodygap}\\
 \frac{\D{x}_j}{\dt}&=-E-\frac{\D {h}_j}{\dt}\frac{{x}_j}{{h}_j}.
 \label{eqn:manybodymass}
\end{align} 
These equations are equivalent to those describing the evolution of a single gap, \eqref{eqn:EOMpair-h}--\eqref{eqn:EOMpair-x}, in the limit of $h\ll1$, but with an effective spring stiffness $\kecjN$. Hence the many-body nature of the cluster problem enters  only through the effective spring stiffness $\kecjN$, which varies through the cluster from relatively soft springs at the edge to relatively stiff springs in the centre. Within the cluster the presence of nearby elements increases the spring force acting to open an individual gap. This is reminiscent of the  `collaborative stiffening' that occurs when elastic plates are stuck together by surface tension: clusters interact just as individual elements do, albeit with a larger effective stiffness that comes from having to bend many plates \cite[][]{Bico2004}. Collaborative stiffening here corresponds to a multi-block cluster whose springs have total stiffness $\sim N\kec$ and typical displacement proportional to $N$. Hence the typical force in the cluster $\sim N^2\kec$. In the two-body problem the displacement close to contact is unity and so, to achieve the same force, the effective stiffness satisfies ${\cal K}\sim N^2\kec$, as seen in \eqref{eqn:Fclust} and \eqref{eqn:manybodygap}.

In \S\ref{phase} we showed for the two-body problem that if $\kec<\keccrit(E)$ the gap thickness will shrink to zero, and the blocks stick; otherwise the blocks will separate and return to their original positions. From the similarity of \eqref{eqn:manybodygap}--\eqref{eqn:manybodymass} to \eqref{eqn:EOMpair-h}--\eqref{eqn:EOMpair-x} at small gap thicknesses, it seems reasonable to expect that, for a cluster containing $\Nblock$ blocks to collapse completely to a final stuck cluster, the effective stiffness $\kecjN$ must be less than $\keccrit(E)$ for each of the gaps within that cluster, i.e.
\beq
\max_{j}\{\kecjN\}<\keccrit(E).
\label{eqn:maxkec}
\eeq  (We assume that any differences between the systems at large gap thicknesses have no significant effect on $\keccrit(E)$.)

Noting that $\max_j\{\kecjN\}=\kec \Nblock^2/4$, we can then restate the criterion in \eqref{eqn:maxkec} as collapse of an $\Nblock$-block cluster (for $\Nblock\geqslant2$) can only occur if \beq
\Nblock<\Nmax= \left(\frac{4\keccrit(E)}{\kec}\right)^{1/2}.
\label{eqn:NmaxCluster}
\eeq Thus if $\Nmax<2$, which occurs at sufficiently large spring stiffnesses,  all blocks end up separate (in a trivial $\Nblock=1$ cluster). More generally,  \eqref{eqn:NmaxCluster} is a prediction for the maximum number of blocks in a cluster as a function of the evaporation rate $E$ and spring stiffness $\kec$. The only assumption used in its derivation was that whether a cluster collapses, and the blocks stick, or splits up should be determined by the behaviour of the system with small gap thicknesses, $h_j\ll1$.

\begin{figure}
\centering
\includegraphics[width=13cm]{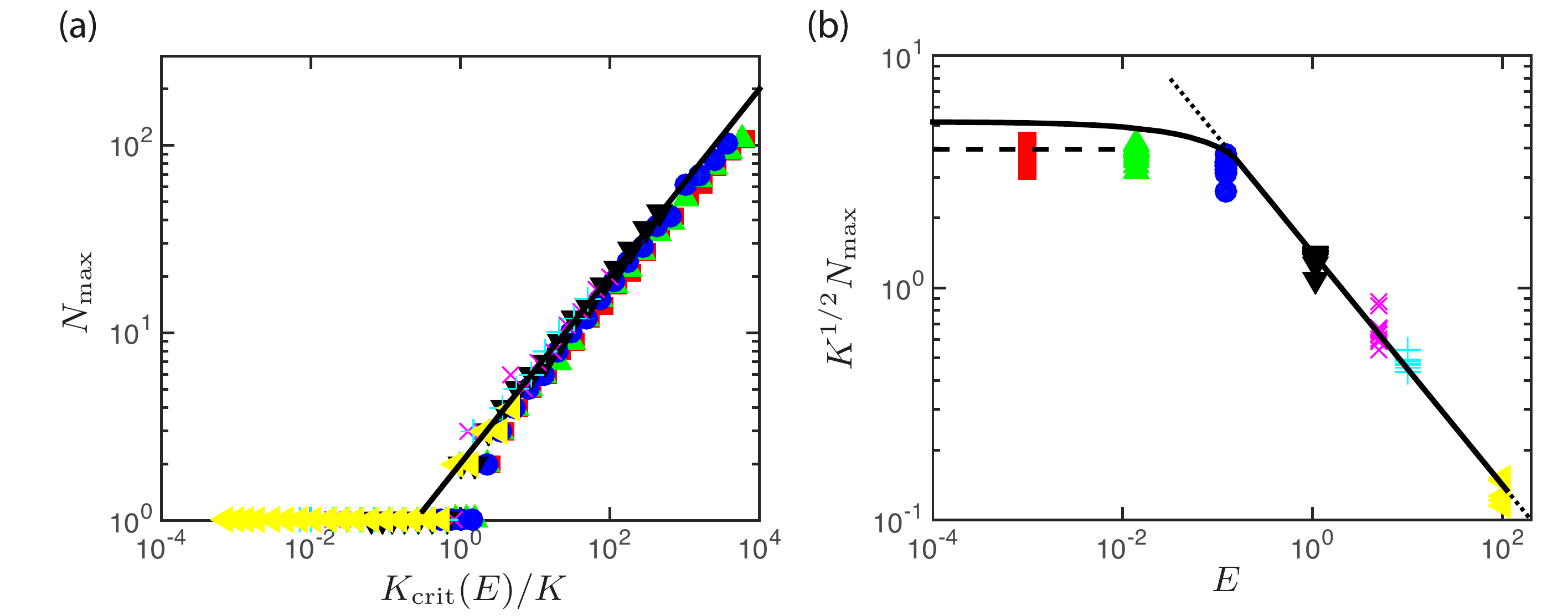}
\caption{ Rescaling of the data in figure \ref{fig:2BodyRegimeDiag}({\it b}) to test the theoretical prediction \eqref{eqn:NmaxCluster} for the maximum number of blocks in a cluster. (a) The maximum number of blocks in any cluster, $\Nmax$, as a function of  $\keccrit(E)/\kec$ (points); equation \eqref{eqn:NmaxCluster} suggests that, provided $\Nmax>1$, this should follow the behaviour $y=2x^{1/2}$ (solid line). (b) An alternative rescaling showing the dependence on the evaporation rate, $E$, directly for data with $\Nmax\geqslant2$. Here, the solid curve shows $2\keccrit(E)^{1/2}$, the dashed horizontal line shows $\Nmax\kec^{1/2}=2^{7/2}\pi/9\approx3.95$ \cite[predicted by][in the limit $E=0$]{Singh2014} and the dotted line shows $\Nmax\kec^{1/2}=(2/E)^{1/2}$, valid for $E\gtrsim0.15$. In both panels, colours and symbols correspond to results obtained at different evaporation rates: $E=10^{-3}$ (red squares), $E\approx10^{-2}$ (green triangles), $E\approx0.1$ (blue circles), $E\approx1$ (black inverted triangles), $E=5$ (magenta `$\times$'), $E=10$ (cyan `$+$') and $E=10^2$ (yellow side triangles).}
\label{fig:NumRes}
\end{figure}

\subsubsection{Reassessment of numerical results}

Figure \ref{fig:NumRes} replots data for the maximum number of  blocks in any cluster from figure \ref{fig:2BodyRegimeDiag}({\it b}), at seven different evaporation rates $E$ and with a variety of spring stiffnesses $K$ (i.e.~the values of $\Nmax$ along seven different horizontal lines in figure \ref{fig:2BodyRegimeDiag}{\it b}). In figure \ref{fig:NumRes}, this data has been rescaled based on the analytical prediction  \eqref{eqn:NmaxCluster}: figure \ref{fig:NumRes}({\it a}) highlights how the raw values of $\Nmax$ vary with $\keccrit(E)/\kec$ while figure \ref{fig:NumRes}({\it b}) highlights the dependence on $E$.

We note first that both rescalings  in figure \ref{fig:NumRes} lead to reasonable collapses of the data from a two-dimensional parameter space (figure \ref{fig:2BodyRegimeDiag}{\it b}) onto a single master curve. Since both of these rescalings follow naturally from \eqref{eqn:NmaxCluster}, we conclude that the reduction from the many-body problem to the two-body problem yields useful insight. However, we also note that for small evaporation rates, $E\ll1$, the maximum cluster size $\Nmax$ lies somewhat below that predicted on the basis of \eqref{eqn:NmaxCluster}. Instead, it seems that for $E\ll1$ the results are closer to the prediction from  the no-evaporation problem, $E=0$,  \cite[][]{Singh2014} where
\beq
\Nmax\kec^{1/2}=\frac{2^{7/2}\pi}{9}\approx3.95.
\label{eqn:E0Nmax}
\eeq Equation \eqref{eqn:E0Nmax} was obtained from an analysis of  front propagation in an unstable medium, which was found to leave behind clusters of this well-defined size. Though larger clusters would, in principle, be able to collapse and not bifurcate, they are apparently not formed by the initial front-propagation mechanism. For all values of $E$, we find that clusters do not reach the maximum size predicted by the linear instability of a nearly uniform state without evaporation, which in this notation reads $\Nmax\kec^{1/2}=2\pi$ \cite[][]{Singh2014}.

\section{Discussion}

We have studied the effect of evaporation on a simple model of elastocapillary aggregation, which revealed two important differences from the same model in the absence of evaporation. Firstly,  evaporation acts to bring the elements together more quickly and leads to contact within a finite time. Secondly, the maximum number of elements in a final cluster decreases with increasing evaporation rate.

The first of these observations is important for applications since it suggests that evaporation can drive objects  together more quickly than would be the case without evaporation, thereby increasing the risk of, for example, MEMS stiction \cite[][]{Tanaka1993}. However, the practical consequences may be (partially) mitigated by the second observation: if high evaporation rates cause fewer elements to come into contact, then those elements that do come into contact are  less likely to be deformed enough to fracture (which is one of the main manufacturing concerns, see figure \ref{fig:setup}{\it a}).

Our analysis of the mathematical model allowed us to reduce the problem of a cluster that is close to collapsed (i.e.~small gap thicknesses) to a two-body problem with an effective spring stiffness. If a cluster starts to form then this effective stiffness is smallest at the edges and largest at the centre of the cluster; thus, if the cluster breaks up (because it is too large) then we expect a void to form in the centre. This is similar to observations of microscopic pillars made up of many carbon nanotubes: if the diameter of the initial pillar (cluster) is small enough then the entire pillar will collapse during evaporation; if not, then voids form in the centre \cite[][]{deVolder2013a}.

We have shown that the number of elements in a cluster is controlled by the dynamics of evaporation. As such, the final pattern observed depends on the dynamics of pattern formation and not simply which pattern is the global energy minimizer, as is often assumed. However, we have also seen  that the effect of evaporation only becomes significant once  $E\gtrsim0.1$, i.e.~when the evaporation dynamics  occur on a time scale comparable to, or faster than, that of the lubrication-type dynamics itself. From \eqref{eqn:Edefn} the dimensionless evaporation rate $E=2\Cae(x_0/w)$, where $\Cae=\mu e/\gamma$ is an evaporative capillary number. Written in this way, we see that $E$ is the product of a property of the liquid filling the gaps ($\Cae$) and the initial aspect ratio of the liquid gaps ($x_0/w$). For the evaporation of pure, still water at room temperature and ambient humidity, $e\approx60\mathrm{~nm/s}$ \cite[][]{Li2012,Routh2013}, $\mu=10^{-3}\mathrm{~Pa\,s}$, $\gamma=72\mathrm{~mN/m}$ so that $\Cae\approx 10^{-9}$. Other liquids evaporate more quickly and hence may have a larger value of $\Cae$. For example, HFE-7100 (3M) has an evaporation rate $e\approx 20\mathrm{~\mu m/s}$ \cite[see][for example]{Lyulin2013,Machrafi2013}, $\gamma=13.6\mathrm{~mN/m}$ and $\mu=6.1\times10^{-4}\mathrm{~Pa\,s}$ so that $\Cae\approx10^{-5}$. Such small values of the evaporative capillary number mean that to obtain $E\gtrsim0.1$ would require $x_0/w\gtrsim 10^4$. While this is at least consistent with the requirement  $x_0/w\gg1$ for lubrication theory to be valid, it is unlikely ever to be the case for the micron-scale plates and pillars often considered experimentally \cite[see][for example]{Wei2015}. However, such large aspect ratios are possible for nanotube forests where the typical length of a nanotube is on the order of $100\mathrm{~\mu m}$ but the radius, and hence the typical separation, is $O(10\mathrm{~nm})$ \cite[][]{Chakrapani2004}.

The geometrical enhancement of the evaporation rate suggests that our prediction that clusters typically contain fewer elements as the evaporation rate increases may be observable in sufficiently dense `forests'. Indeed, from their experiments on the formation of cells in carbon nanotube forests (see figure \ref{fig:setup}{\it b}), \cite{Chakrapani2004} report that ``[a faster rate of evaporation] favors crack growth and results in a decrease in the average cell width" (i.e.~fewer elements form in the clusters between cracks). However, they do not present quantitative data to support this comment. 

We hope that the theoretical results presented here may motivate experimental efforts to better understand how evaporation rate influences pattern formation in these systems. At the same time, there are a number of improvements that should be made to the  model developed here. For example, experiments are bound to take place in a three-dimensional setting where the force from surface tension will require a more detailed calculation \cite[][]{Wei2015} and the resistive forces from the lubricating flow may be reduced since liquid can flow horizontally, as well as vertically.

Even in situations where the delicate balance between evaporation and lubrication dynamics discussed here is not appropriate,  other, slower timescales exist with which evaporation can more readily compete. For example the viscoelastic relaxation time of the elements themselves may be important \cite[][]{Wei2015}. We expect that the techniques developed in this paper may be applied, with suitable modifications, to understand the interplay between evaporation and dynamics in such systems.

\section*{Acknowledgements}
KS and DV wish to acknowledge the support of the King Abdullah University of Science and Technology (KAUST; Award No. KUK-C1-013-04), and the John Fell Oxford University Press (OUP) Research Fund. 

\appendix

\section{Derivation of model}
\label{sec:appDeriv}

In this Appendix, we outline the derivation of the expression for the force exerted by the $j$-th liquid gap on each of its neighbouring blocks, \eqref{eqn:ND_FP}. More details are given by \cite{Singh2014}.
The well-known lubrication equation, together with the local conservation of mass leads to a relationship between the pressure profile within the gap, $p_j(x)$, and the rate at which the gap thickness, $h_j$, is changing with time. In particular,
\beq
\frac{\upd h_j}{\upd t}=\frac{h_j^3}{12\mu}\frac{\partial^2 p_j}{\partial x^2}.
\eeq This equation for $p_j$ is to be solved subject to the boundary conditions of no flux through the base, i.e.~ $\partial p_j/\partial x|_{x=0}=0$, and that the pressure just beneath the meniscus is equal to the capillary pressure, i.e.~$p_j(x_m^-)=-2\gamma/h_j$. We find that
\beq
p_j(x)=\frac{6\mu }{h_j^3}\frac{\upd h_j}{\upd t}(x^2-x_j^2)-\frac{2\gamma}{h_j}.
\eeq

The force on each of the blocks that bound the $j$-th gap is then given by
\beq
F_j=-\int_0^{x_j}p_j~\upd x=\frac{4\mu x_j^3 }{h_j^3}\frac{\upd h_j}{\upd t}+\frac{2\gamma x_j}{h_j},
\eeq which is the dimensional version of \eqref{eqn:ND_FP}.

\section{Phase-plane analysis to determine $\keccrit(E)$}
\label{sec:appPP}

For $0<\KE<\frac12$ the phase plane of \eqref{eqn:EOMpair-hsc}--\eqref{eqn:EOMpair-xi} is topologically equivalent to figure 2({\it a}): trajectories generically tend either to the stable node  $(0,1)$ (separated blocks) or to the stable node  $(\xip,0)$ (stuck blocks). The dividing trajectory between these outcomes is the stable manifold of the saddle point  $(\xi_s,0)$, which is given by $\xi\sim \xi_s+2h\xi_s/(\xi_s-3)+O(h^2)$ as $h\to0$. Starting with this limit, we can integrate back along the stable manifold to $h=1$, which defines a point $(\keccrit^{-1},1)$. Then if $\kec\leqslant \keccrit$ an initial condition $(\kec^{-1},1)$ leads to a stuck state.

As $\KE\nearrow \frac12$ the points $(\xi_s,0)$ and $(\xip,0)$ merge into a saddle-node at $(2,0)$ and $\keccrit=3.30+O(1-2\KE)^{1/2}$. For $\KE>\frac12$ all trajectories lead to the separated state.

At $\KE=0$ the (nontrivial) nullclines of \eqref{eqn:EOMpair-hsc}--\eqref{eqn:EOMpair-xi} coincide as a line of fixed points $h+\xi=1$, corresponding to a mechanical equilibrium $\kec(1-h)=x/h$ between the spring and capillary forces in the absence of evaporation. Away from this equilibrium, the trajectories satisfy $h^2\xi=\textrm{const.}$, corresponding to mass conservation $hx=\textrm{const.}$.

To consider the useful limit of slow but non-zero evaporation,  
we let $\epsilon=\KE\ll1$ and note that the stable manifold of the saddle point, $\xi=1+\epsilon/2-h(1+3\epsilon/4)+O(\epsilon^2,h^2)$, starts almost parallel to the two nearly coincident nullclines. This motivates setting $\eta=(h+\xi-1)/\epsilon$ so that \eqref{eqn:EOMpair-hsc}--\eqref{eqn:EOMpair-xi} become
\beq
\frac{\upd \xi}{\upd s}=\epsilon \xi\left[2\eta-\xi^2\right],\qquad
\frac{\upd \eta}{\upd s}=(3\xi-1)\eta-\xi^3-\epsilon \eta^2.\label{eqn:etaxi}
\eeq
Since $\epsilon\ll1$, we expect the evolution of $\xi$ to be slow and hence, integrating backwards, $\eta$ collapses onto a `slow' manifold given by ${\upd \eta}/{\upd s}=O(\epsilon)$, or
\beq
\eta\sim\frac{\xi^3}{3\xi-1},
\label{eqn:slow}
\eeq over the range $1/3<\xi\leqslant 1$. As $\xi$ approaches $1/3$, $\eta$ diverges and we need to rescale the equations to understand how the stable manifold leaves the $O(\epsilon)$ neighbourhood of $h+\xi=1$.  

A suitable rescaling is to set $x=3(3\xi-1)/2\epsilon^{1/3}$, $y=9\eta \epsilon^{1/3}=9(\xi+h-1)/\epsilon^{2/3}$. At leading order \eqref{eqn:etaxi} becomes
\beq
\frac{3}{\epsilon^{1/3}}\frac{\upd x}{\upd s}=y,\quad \frac{3}{\epsilon^{1/3}}\frac{\upd y}{\upd s}=2xy-1, \mbox{~ and thus~} \frac{\upd y}{\upd x}=2x-\frac1y.
\label{eqn:xy}
\eeq
For the solution to \eqref{eqn:xy} to match onto \eqref{eqn:slow}, we require $y\sim 1/2x$ as $x\to \infty$. This condition specifies a unique solution to \eqref{eqn:xy}, given implicitly by $x=-{\rm Ai}'(z)/{\rm Ai}(z)$, where $\rm Ai$ denotes the Airy function and $z=x^2-y$. As $x\to-\infty$, $y\sim x^2-z_0$, where $z_0=-2.338$ is the first zero of ${\rm Ai}(z)$. This asymptotic behaviour matches onto the trajectory $h^2\xi=(4/27)(1-\epsilon^{2/3}z_0/3)+O(\epsilon)$, which continues to $(\keccrit^{-1},1)$ . Hence $\keccrit\sim (27/4)(1+\epsilon^{2/3}z_0/3)\sim (27/4)-18.8E^{2/3}$. The limiting value $\keccrit= 27/4$ for $E=0$ was obtained by \cite{Singh2014}.


\end{document}